\documentclass[11pt]{article}
\pdfoutput=1
\usepackage{moriond}
\usepackage{amssymb}
\usepackage{cite}
\usepackage{epstopdf}

\def\PRD{{\em Phys. Rev.} D}

\def\be{\begin{equation}}
\def\ee{\end{equation}}
\def\bea{\begin{eqnarray}}
\def\eea{\end{eqnarray}}

\def\pt {{$p_T$}}

\def\ttbar {{$t \bar{t}$}}
\def\Z {{$Z/\gamma^*$}}
\def\MET {{\not\! E_T}}

\newcommand{\Photo}{\includegraphics[height=35mm]{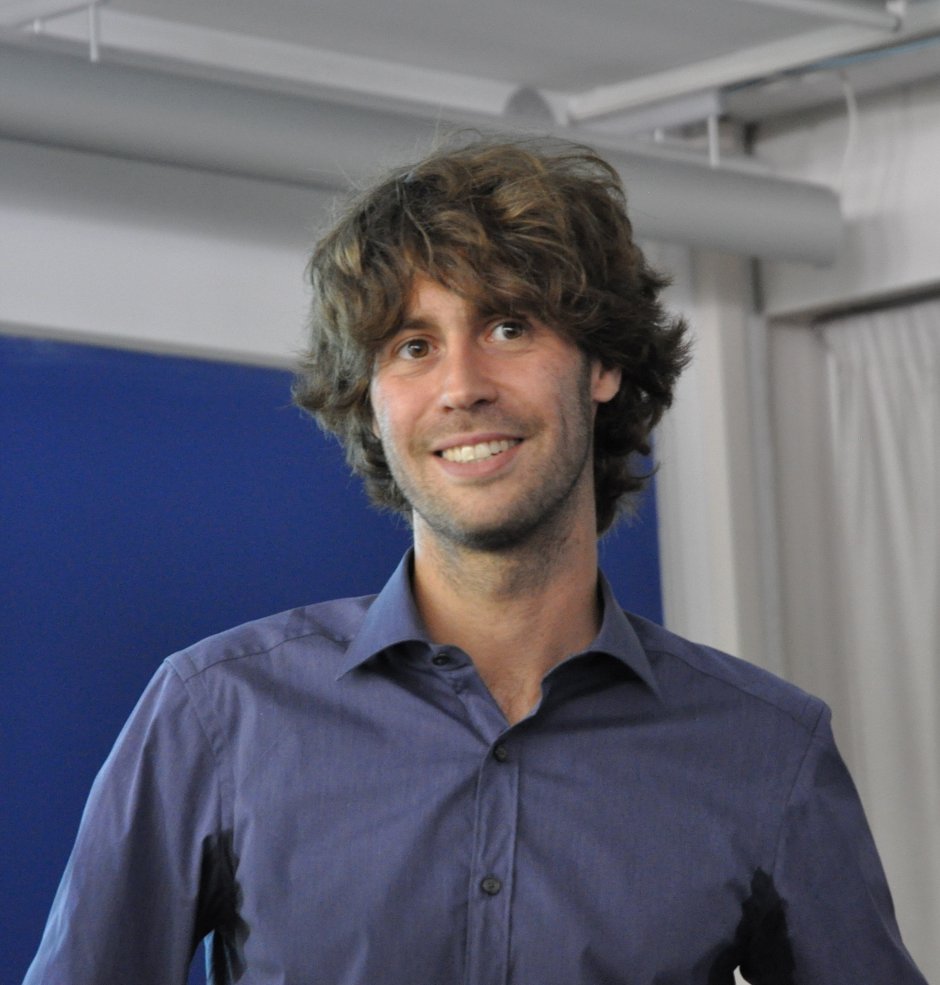}}

\begin{document}
\vspace*{4cm}
\title{W/Z+Jets and W/Z+HF Production at the Tevatron}

\author{Stefano Camarda\footnote{formerly IFAE - Barcelona}}
\address{DESY - Notkestra{\ss}e 85
D-22607 Hamburg Germany\\
On behalf of the CDF and DO Collaborations.}

\maketitle\abstracts{
The CDF and D0 collaborations performed a comprehensive study of 
the production of vector bosons, W and Z, in association with
energetic jets. Understanding the W/Z + jets
and W/Z + c, b-jets processes is of paramount importance for
the top quark physics, for the Higgs boson measurements, and for many new physics
searches. In this contribution the most recent measurements of the associated
production of jets and vector bosons in Run II at the Tevatron are
presented. The measurements are compared to different perturbative QCD predictions
and to several Monte Carlo generators.}

\section{Introduction}
The study of the production of electroweak bosons in association with
jets of hadrons constitutes a fundamental item in the high-$\rm p_T$
physics program at the Tevatron. Vector bosons plus jets final states are a
major background to many interesting physics processes like single and
pair top quarks production, Higgs, and super-symmetry. Precise
measurements of W/Z + jets production provide a stringent test of perturbative QCD
predictions~\cite{QCD, MCFM, Blackhat, Loopsim, QCDEW} at high $Q^2$, and offer the possibility to
validate Monte Carlo simulation tools~\cite{Alpgen, Powheg, Pythia, Herwig, Sherpa, HEJ}.
The latest vector boson plus jets results at the Tevatron are reviewed and discussed.

\section{W/Z + jets measurements}
\begin{figure}[htbp]
  \begin{center}
    \includegraphics[width=0.9\textwidth]{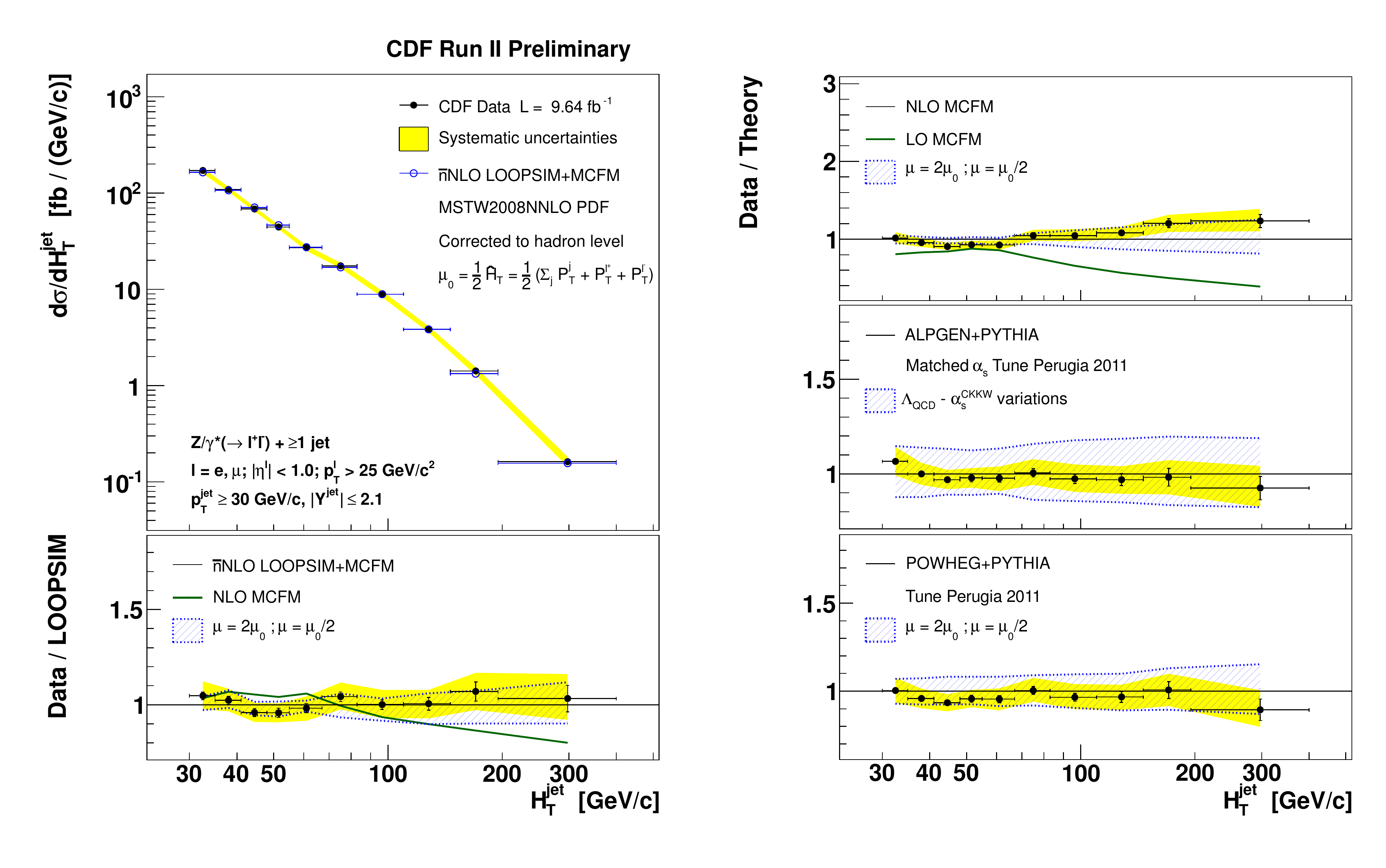}
  \end{center}

    \caption{Measured differential cross sections 
as a function of $H_T^{\textrm{jet}}$ in \Z{} $\rm+ \geqslant 1~jet$
events. Data (black dots) are compared to \textsc{loopsim} prediction (open circles). The shaded bands show
the total systematic uncertainty, except for the 5.8$\%$ luminosity uncertainty, 
the blue area represents simultaneous variation of renormalization and factorization scales.}
	\label{fig:zjets}
\end{figure}

The CDF experiment recently measured Z + jets production
cross sections with the full Tevatron run II dataset, corresponding to $9.64~\textrm{fb}^{-1}$ of
integrated luminosity~\cite{Camarda:2012yha}. Differential cross sections as a function of several variables have been measured, including jet
\pt{}, jet rapidity and jet multiplicity, angular variables like di-jet $\Delta \phi$ and $\Delta y$, and $H_T^{\textrm{jet}} = \sum p_T^{\textrm{jet}}$.
Events are required to have two electrons or muons with a
reconstructed invariant mass in the range $66 \leqslant M_{Z}
\leqslant 116 \rm ~GeV/c^{2}$ around the Z boson mass.
Jets are clustered with the Midpoint algorithm~\cite{midpoint} in a
cone radius of $0.7$, and are required to have $p_T \geqslant
30~ \rm GeV/c$ and $|y| \leqslant 2.1$. 
The background estimation is
performed both with data-driven and Monte Carlo techniques, the QCD and W+jet
backgrounds are estimated from data,
other backgrounds contributions like \ttbar{}, diboson and $Z/\gamma^* \rightarrow \tau^+ \tau^-$ are estimated
from Monte Carlo simulation.
The cross sections are unfolded back to the particle level accounting for
acceptance and smearing effects employing \textsc{alpgen+pythia} Monte Carlo.
The measured cross sections are compared to predictions from the Monte Carlo event generators \textsc{alpgen+pythia}\cite{Alpgen},
\textsc{powheg+pythia}\cite{Powheg}, and to the fixed order perturbative QCD predictions \textsc{mcfm}\cite{MCFM}, \textsc{blackhat+sherpa}\cite{Blackhat}, \textsc{loopsim+mcfm}\cite{Loopsim} and
to a fixed order prediction including NLO EW corrections\cite{QCDEW}.
Fixed order predictions include parton-to-particle correction factors that account for the non-perturbative underlying event and
fragmentation effects, estimated with \textsc{alpgen+pythia} Monte Carlo simulation.
Figure \ref{fig:zjets} shows the measured cross section
as a function of $H_T^{\textrm{jet}}$ in \Z{} $\rm+ \geqslant 1~jet$ events.

\begin{figure}[htbp]
  \begin{center}
    \includegraphics[width=0.45\textwidth]{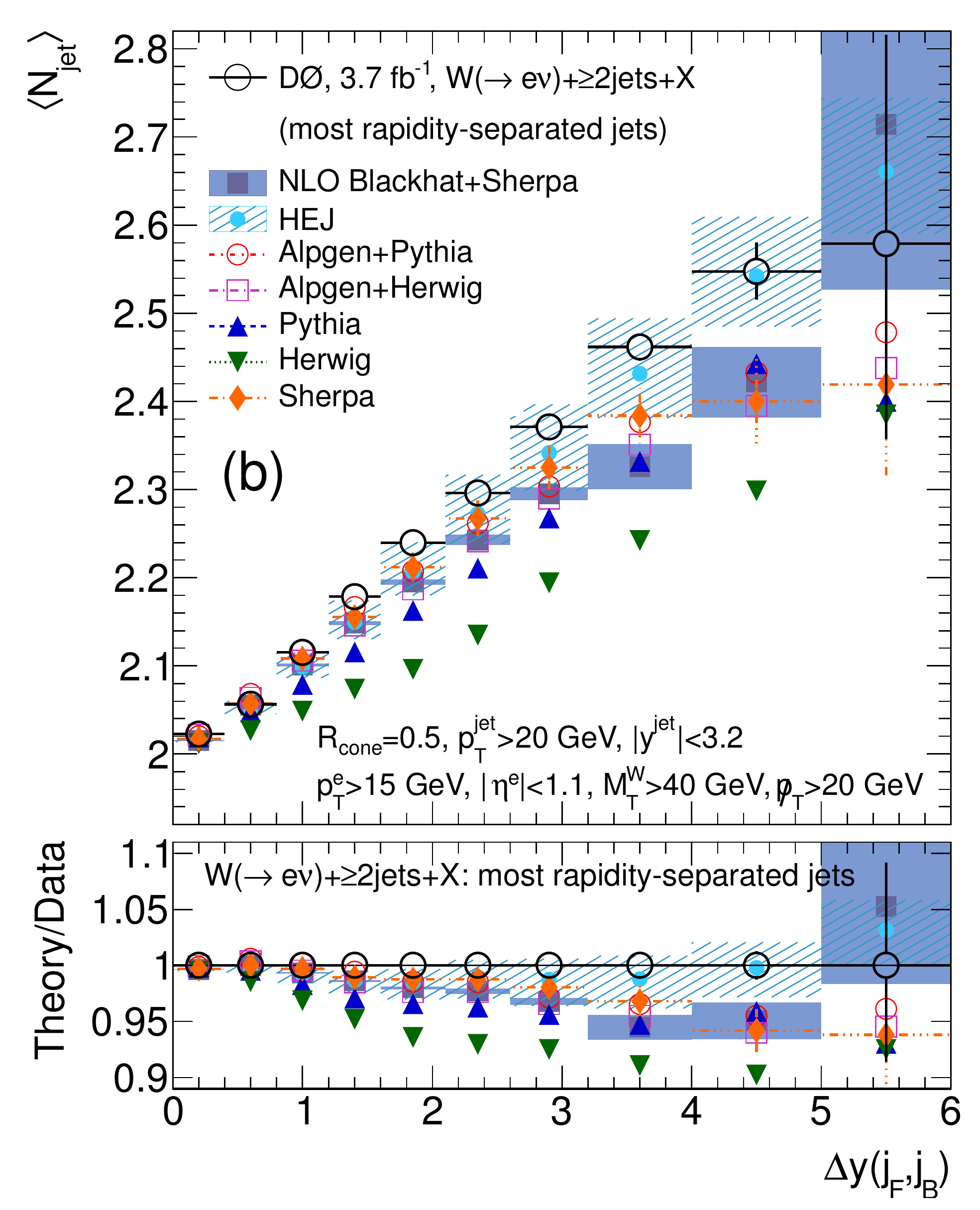}
    \includegraphics[width=0.45\textwidth]{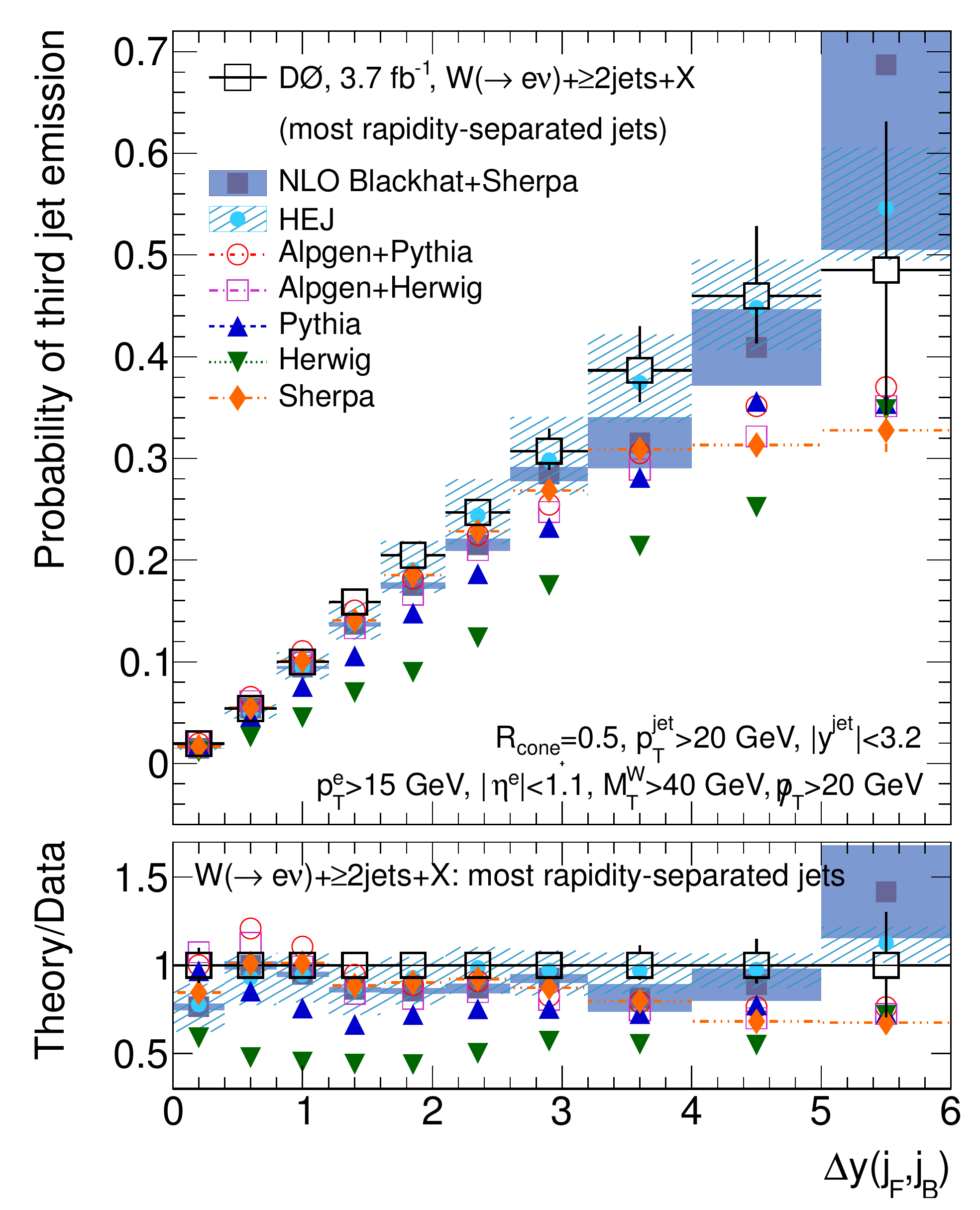}
    \caption{Measured average number of jets (left) and probability of third jet emission (right)
as a function of the $\Delta y$ between the most rapidity-separated jets, in events with $W \rightarrow e \nu \rm+ \geqslant 2~jets$.
Data (open black dots) are compared to several Monte Carlo generators and predictions. The lower pane shows the theory/data ratio.}
	\label{fig:D0wjets}
  \end{center}
\end{figure}

A new measurement of $W \rightarrow e \nu$ + jets production cross section has been performed
with the D0 experiment with $3.7~\rm fb^{-1}$ of integrated luminosity~\cite{D0Wjets}, 
including a comprehensive study of several kinematics variables.
Events are selected with a reconstructed electron of $p_T
\geqslant 15$ GeV/c and $|\eta| \leqslant 1.1$, the transverse mass of
the W, reconstructed with the electron and $\MET$, is required to be
$M_{T}^{W} \geqslant 40~ \rm GeV/c^2$, jets are reconstructed with the Midpoint algorithm 
in a radius $R=0.5$ and required to have $p_T \geqslant 20$ GeV/c and $|y| \leqslant 3.2$.
Data are compared to several Monte Carlo generators, \textsc{alpgen+pythia}, \textsc{alpgen+herwig}\cite{Alpgen}, 
\textsc{pythia}\cite{Pythia}, \textsc{herwig}\cite{Herwig} and \textsc{sherpa}\cite{Sherpa}, to perturbative NLO QCD
predictions from \textsc{blackhat+sherpa}\cite{Blackhat}, and to the all order resummed prediction \textsc{hej}\cite{HEJ}.
Figure \ref{fig:D0wjets} shows the average number of jets and the probability of third jet emission as a function of the $\Delta y$ 
between the most rapidity-separated jets, in events with $W \rightarrow e \nu \rm+ \geqslant 2~jets$.

\section{W/Z + heavy flavor jets production}
The measurement of vector boson production with associated heavy
flavor jets provides an important test of perturbative QCD predictions, and
can be used to improve the determination of PDF.
Understanding these processes is also critical for the measurement
of Higgs boson production in association with a W or Z and in
the search for SUSY.

The W + charm production cross section has been measured by CDF
with $4.3~\rm fb^{-1}$ of integrated luminosity\cite{WcharmCDF}.
Charm jets are identified with an algorithm which identifies soft leptons
coming from the semileptonic decay of the charm.
The measurement exploits the charge correlation between the soft lepton and the lepton
coming from the leptonic decay of the W to reduce the background contamination.
The measured cross section of $\rm 13.6 \pm 2.2(stat) ^{+2.3}_{-1.9} (syst) \pm 1.1 (lum)$ pb
is in good agreement with the NLO prediction of $11.4 \pm 1.3$ pb from MCFM.

The W + b-jet cross section has been measured by D0
with $6~ \rm fb^{-1}$ of integrated luminosity\cite{D0Wbjet}.
$W \rightarrow e \nu$ and $W \rightarrow \mu \nu$ decay channels are combined, 
and a multivariate technique is used to identify b-jets.
The measured cross section of $1.05 \pm 0.12$ pb is in good
agreement with the perturbative QCD NLO prediction from
MCFM of $1.34 ^{+0.41}_{-0.34}$ pb, and consistent with predictions from the Monte
Carlo generators \textsc{sherpa} ($1.08$ pb) and \textsc{madgraph} ($1.44$ pb).

The CDF and D0 collaborations recently measured differential cross sections of Z + b-jet production
with the full Tevatron run II dataset\cite{CDFZbjet, D0Zbjet}, with integrated luminosities corresponding
to $9.7~\rm fb^{-1}$ and $9.13~\rm fb^{-1}$ respectively.
\begin{figure}[htbp]
  \begin{center}
    \includegraphics[width=0.4\textwidth]{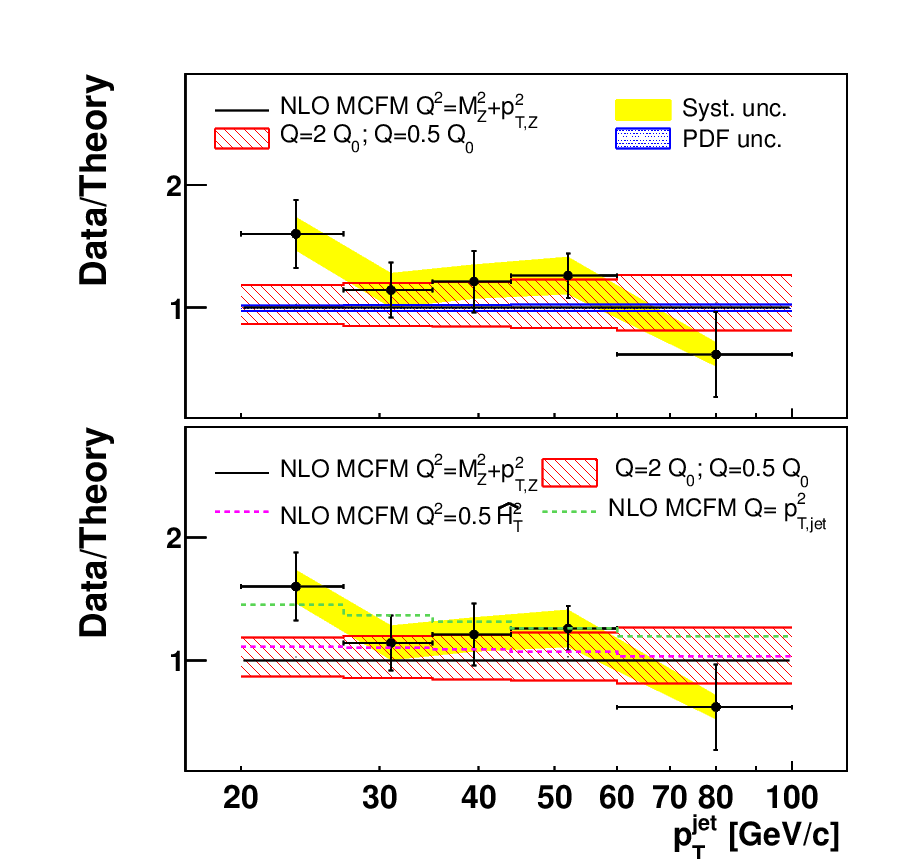}
    \includegraphics[width=0.35\textwidth]{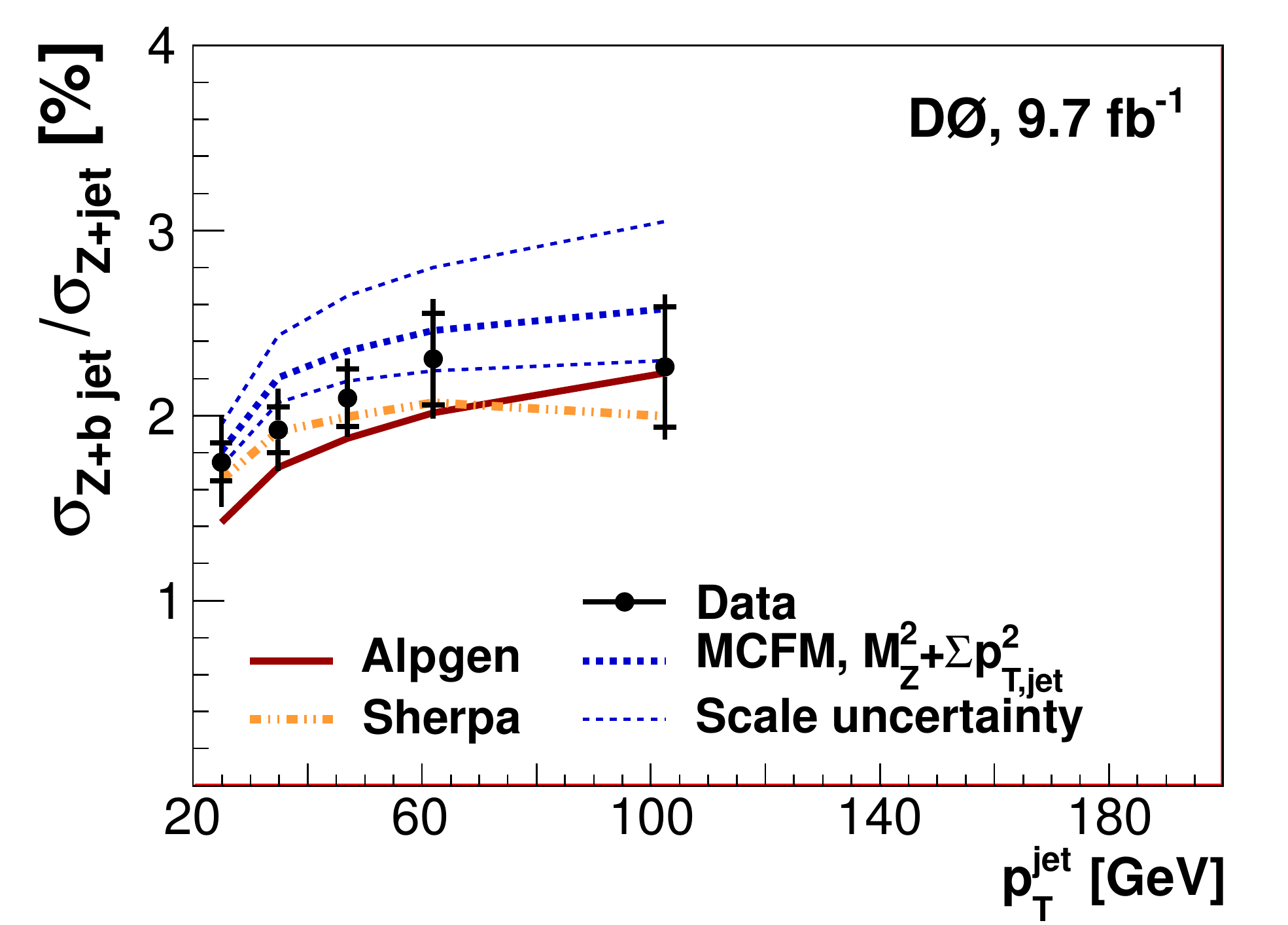}
    \caption{Measured Z + b-jet differential cross sections as a function of b-jet $p_T$ with the CDF (left) and D0 (right) detectors.
Data (black dots) are compared to MCFM NLO prediction and Monte Carlo generators.}
	\label{fig:Zbjet}
  \end{center}
\end{figure}
Figure \ref{fig:Zbjet} shows the measured cross sections unfolded to particle level and 
compared to NLO perturbative QCD predictions from MCFM and Monte Carlo generators \textsc{sherpa}
and \textsc{alpgen+pythia}.
The measured cross sections are in reasonable agreement with theory, within large experimental
and theoretical uncertainties.

\section*{Summary}
W/Z + jets and W/Z + heavy flavor measurements belong to the Tevatron legacy.
All the measurements are in general good agreement with the perturbative NLO QCD
predictions, in the tail of some distributions like $H_T^{\rm jet}$ and di-jet $\Delta y$ the inclusion
of higher order corrections improves the agreement between data and theory.
Detailed studies of differential distributions in W/Z + jets and W/Z + heavy flavour production provide an important test of
the different Monte Carlo generators and theoretical predictions, and a fundamental validation
of the background modeling of such processes in the search for new physics.

\section*{Acknowledgments}
I am grateful to the ATLAS-DESY group for supporting my participation.

\end{document}